\documentclass[aps,prb,twocolumn,showpacs,superscriptaddress,floatfix,10pt]{revtex4-1}

\usepackage{graphicx}
\usepackage{amsmath}
\usepackage{epsfig}
\usepackage{helvet}
\usepackage{amssymb}
\usepackage{framed}
\usepackage{enumitem}
\usepackage[bookmarksnumbered=true,hypertexnames=false,colorlinks=true,linkcolor=blue,urlcolor=blue,citecolor=blue,anchorcolor=green]{hyperref}

\newcommand{\ket}[1]{\mbox{$| #1 \rangle$}}
\newcommand{\braket}[2]{\mbox{$\langle #1 | #2 \rangle$}}

\newcommand{\frw}[1]{$\overset{\lower0.5em\hbox{$\smash{\scriptscriptstyle\smile}$}} #1$}

\def\T{{\mathcal T}}

\begin{document}

\title{Gauge fixing, canonical forms and optimal truncations \\ in tensor networks with closed loops}
\author{Glen Evenbly}
\affiliation{D\'epartement de Physique and Institut Quantique, Universit\'e de Sherbrooke, Qu\'ebec, Canada
}
\email{glen.evenbly@usherbrooke.ca}
\date{\today}

\begin{abstract}
We describe an approach to fix the gauge degrees of freedom in tensor networks, including those with closed loops, which allows a canonical form for arbitrary tensor networks to be realised. Additionally, a measure for the internal correlations present in a tensor network is proposed, which quantifies the extent of resonances around closed loops in the network. Finally we describe an algorithm for the optimal truncation of an internal index from a tensor network, based upon proper removal of the redundant internal correlations. These results, which offer a unified theoretical framework for the manipulation of tensor networks with closed loops, can be applied to improve existing tensor network methods for the study of many-body systems and may also constitute key algorithmic components of sophisticated new tensor methods.
\end{abstract}

\maketitle

\section{Introduction} 
Tensor network methods\cite{TN1,TN2,TN3} have proven to be exceptionally useful in the study of quantum many-body system and, more recently, have also found a diverse range of applications in areas such as quantum chemistry\cite{Chem1,Chem2}, machine learning\cite{Mach1,Mach2,Mach3}, and holography\cite{Holo1,Holo2,Holo3,Holo4}. In context of many-body systems, tensor networks circumvent the exponentially growth of Hilbert space with system size by allowing a quantum many-body wavefunction to be expressed as a product of many small tensors. As exemplified by White's density matrix renormalization group (DMRG) algorithm \cite{DMRG1,DMRG2,DMRG3}, which is based on matrix product states\cite{MPS1,MPS2} (MPS), and through more newly developed algorithms such as those based on projected entangled pair states\cite{PEPS1,PEPS2,PEPS3} (PEPS) and the multi-scale entanglement renormalization ansatz\cite{MERA1,MERA2,MERA3,MERA4} (MERA), tensor networks can potentially allow many-body systems to be accurately addressed in the thermodynamic limit directly. 

Key to formulation of DMRG, by far the most widely established tensor network method, is the singular value decomposition (SVD), also called the Schmidt decomposition in the context of quantum information theory\cite{Neilson}. The Schmidt decomposition is an integral part of the DMRG algorithm as it allows (i) the gauge degrees of freedom in an MPS to be fixed, in turn leading to the notion of a canonical form for MPS\cite{CN1}, and (ii) the internal indices of MPS to be truncated in an optimal way. Use of the Schmidt decomposition can also be extended to arbitrary loop-free (or acyclic) tensor networks, generically referred to as tree tensor networks\cite{TTN1,TTN2} (TTN), of which MPS is a particular instance. However for tensor networks that contain closed loops, such as PEPS or MERA, the Schmidt decomposition is no longer applicable. Thus there does not exist a well-defined canonical form for networks with closed loops, nor is it easy to truncate internal indices in an optimal manner. These difficulties have been a major stumbling block in the development of tensor network algorithms for quantum systems in $D>1$ dimensions, such as those based on PEPS.

In this manuscript we present, for an arbitrary tensor networks including those with closed loops (also called cyclic networks), (i) a method of fixing the gauge degrees of freedom, related to a generalization of the Schmidt decomposition, (ii) a means of quantifying the extent of \emph{internal correlations} through closed loops and (iii) an algorithm for optimally truncating internal indices, which can potentially remove the redundant internal correlations from the network description. We demonstrate that the proposed method of fixing gauge degrees of freedom has the same uniqueness properties as that of the Schmidt decomposition, such that it may be used to define a canonical form for arbitrary tensor networks. As many commonly used tensor network optimization schemes rely on the accurate truncation of network indices, the result (iii) could have substantial application across a variety of different state-of-the-art tensor network algorithms. For instance, in methods for the renormalization of tensor networks\cite{TRG1,TRG2,TRG3,TRG4,TRG5,TRG6,TRG7}, the proper removal of internal correlations from within closed loops was demonstrated to be of key importance with the advent of tensor network renormalization\cite{TNR1,TNR2,TNR3,TNR4} (TNR) and related approaches\cite{TNRc1,TNRc2,TNRc3,TNRc4,TNRc5}. It follows that the proposal for removing internal correlations presented here may also be applied as a core part of a TNR-like numerical method for the simulation of a many-body systems.

This manuscript is organised as follows. First we refresh basic notions of gauge freedom in tensor networks, before introducing the concept of a \emph{bond environment}. We then present an algorithm for fixing the gauge of an index, and argue that it converges to a unique gauge in a general network. The concept of internal correlations in cyclic networks is then discussed, including the proposal of a measure to quantify them. An algorithm for truncation of internal indices is then proposed, and subsequently demonstrated to be effective in the removal of internal correlations from cyclic networks. Finally, we discuss some of the applications of the methods presented.

\section{Tensor networks} 
\begin{figure}[!t!b]
\begin{center}
\includegraphics[width=8.5cm]{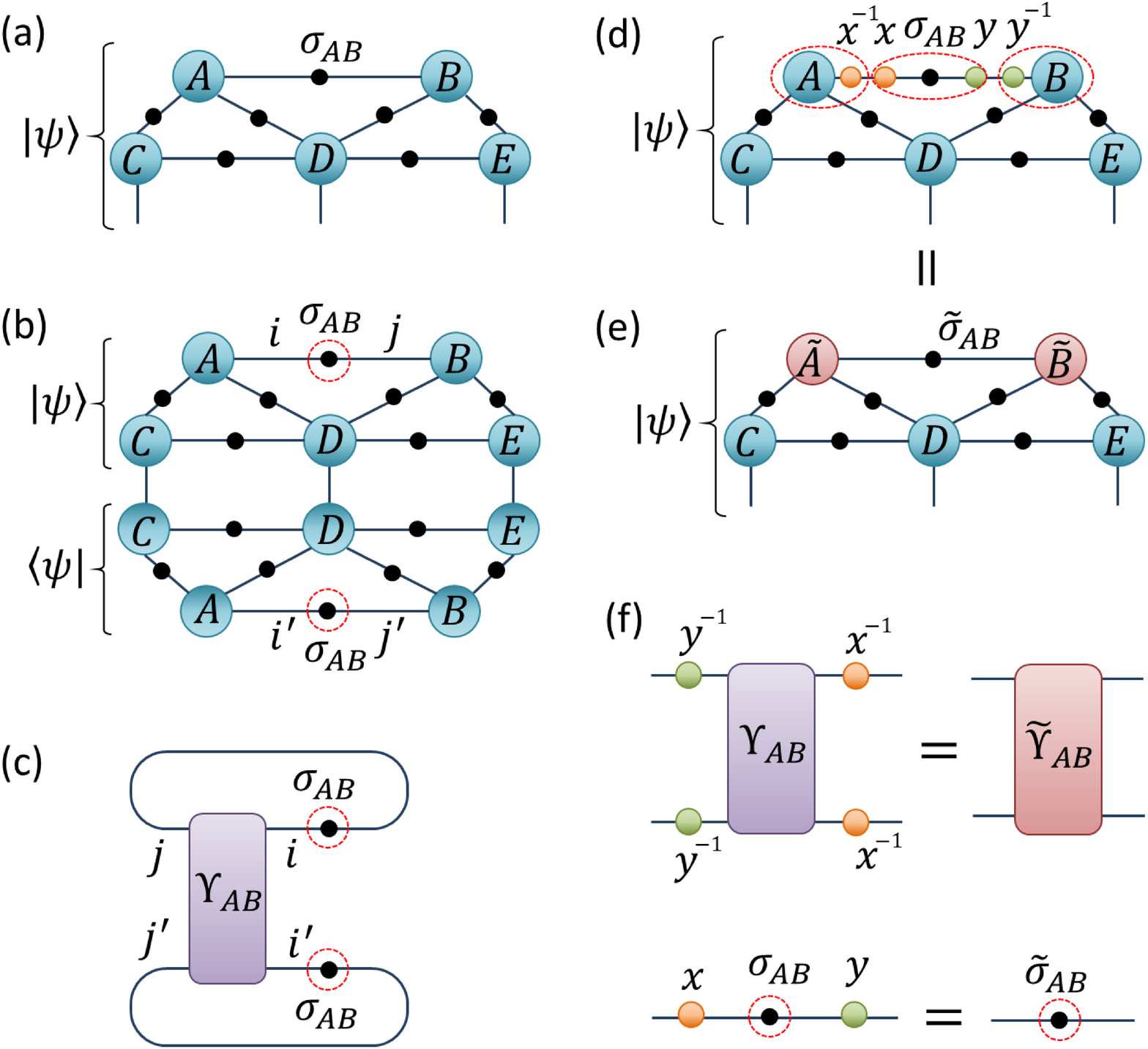}
\caption{(a) Quantum state $\ket{\psi}$ defined from a network of tensors $\{A,B,C,D,E \}$ with bond matrices $\sigma$ sitting between pairs of tensors. (b) Tensor network for $\braket{\psi}{\psi}$. (c) The bond environment $\Upsilon_{AB}$ is defined by contracting $\braket{\psi}{\psi}$ while leaving the index between $A$ and $B$ (and their conjugates) open. (d-e) A change of gauge, which leaves the state $\ket{\psi}$ invariant, is enacted on the index between $A$ and $B$ via matrices $x$ and $y$ together with their inverses. (f) Depiction the new bond environment $\tilde{\Upsilon}_{AB}$ and associated bond matrix ${\tilde \sigma}_{AB}$ from the gauge change in (e).} 
\label{fig:GaugeChange}
\end{center}
\end{figure}

We consider a network $\T$ composed of a set of tensors $\left\{ A,B,C,\ldots \right\}$, as depicted in Fig. \ref{fig:GaugeChange}, and distinguish between \emph{internal indices}, which each connect a pair of tensors within the network, and \emph{external indices} which each only attach to a single tensor. Additionally, for future convenience, we allow the possibility a \emph{bond matrix} $\sigma$ to be situated on each internal index, e.g. where $\sigma_{AB}$ denotes the bond matrix situated between tensors $A$ and $B$. The bond matrices could initially bet set as (trivial) identity matrices, $\sigma = \mathbb{I}$, if desired. To each external index $\alpha$ one associates a Hilbert space $\mathbb{V}_\alpha$ of equal dimension, such that the network $\T$ can be interpreted as describing a quantum state $\ket{\psi}$ on the tensor product space, $\mathbb{V}_\textrm{full} = \left( \otimes \mathbb{V}_\alpha \right)$, where the product is over all external indices. An internal index is called a \emph{bridge} if its removal would split the tensor network into two disconnected components (or equivalently, an index is a bridge if and only if it is not contained in any cycle). It follows that, in a loop-free tensor network (also called an acyclic network), all internal indices are bridges. Recall that there is a gauge freedom on the internal indices of a network $\T$; introducing an arbitrary invertible matrix $x$ and its inverse $x^{-1}$ on an internal index leaves the state $\ket{\psi}$ invariant, but the network representation is changed when absorbing $x$ and $x^{-1}$ into the adjoining tensors as depicted in Fig. \ref{fig:GaugeChange}(d-e).

Before we discuss a means by which to fix the gauge degrees of freedom it is useful to introduce the concept of a \emph{bond environment}. For any internal index the bond environment $ \Upsilon_{i'j'}^{ij}$ is a four index tensor defined through contraction of $\braket{\psi}{\psi}$ while leaving the indices connected to the associated bond matrix $\sigma$ (and its conjugate\cite{real}) open, see Fig. \ref{fig:GaugeChange}(b-c). It follows that the scalar product $\braket{\psi}{\psi}$ is obtained by contracting a bond environment with the two associated bond matrices, 
\begin{equation}
\left\langle {\psi }
 \mathrel{\left | {\vphantom {\psi  \psi }}
 \right. \kern-\nulldelimiterspace}
 {\psi } \right\rangle  = \sum\limits_{i,j,i',j'} {\Upsilon _{i'j'}^{ij}{\sigma _{ij}}{\sigma _{i'j'}}}.
\end{equation}
Bond environments $\Upsilon$ have many useful properties, including:
\begin{enumerate}[label=\textbf{E.\arabic*},ref=E.\arabic*]
\item The bond environment of an index is invariant with respect to choice of gauge on all other internal indices of the network \label{E1}
\item All bond environments are invariant with respect to unitary transfomation acting on the external indices of the network \label{E2} 
\item A bond environment factorizes into a product of two tensors, $\Upsilon_{i'j'}^{ij} = \left(\Upsilon_R \right)^i_{i'} \left(\Upsilon_L \right)^j_{j'}$, if the associated index is a bridge. \label{E3}
\end{enumerate}
It should be clarified that property \ref{E2} refers to invariance with respect a unitary transformation that can act jointly over all the external indices of a network, not only unitary transformations acting singularly on each external index. Notice that property \ref{E2} further implies that the factorization of \ref{E3} occurs if there \emph{exists} a unitary transformation $U$ on the external indices that would allow the index to become a bridge in the transformed network (even if this specific $U$ is unknown), see Sect. \ref{sect:A} of the appendix for further discussion. The concept of a bond environment is key to the results presented in this manuscript: the algorithms that we present for (i) fixing the gauge on an index, for (ii) quantifying the internal correlations though an index and for (iii) truncating the dimension of an index each only require the corresponding bond environment and bond matrix as inputs.

\section{Gauge fixing} If an internal index of a tensor network $\T$ is a bridge, then the gauge freedom can be fixed using by imposing a Schmidt form with respect to that index\cite{TN2, TN3}. This involves choosing the gauge such that (i) the sub-networks on each side of the bridge each represent an orthonormal basis (in the respective Hilbert spaces corresponding to their external indices) and (ii) the bond matrix $\sigma$ is diagonal and positive, $\sigma_{ij} = \delta_{ij} s_i$ with $s_i$ the Schmidt coefficients, which are ordered $s_i \ge s_{i+1}$. While fixing an index in Schmidt form removes most of the gauge freedom, some freedom can still remain. Specifically, if two or more of the Schmidt coefficients are exactly degenerate, then there remains a unitary gauge freedom within the degenerate subspace. There is also a \emph{phase ambiguity}; the Schmidt form is still retained under a gauge change (as described in Fig. \ref{fig:GaugeChange}(d-e)) with $x$ and $y$ as diagonal matrices with entries of unit magnitude, i.e. $x_{ij} = y_{ij} = \delta_{ij} \exp(\mathbf{i} \theta_i)$ for some real angles $\theta_i \in [-\pi, \pi]$ and with $\mathbf{i}$ the complex unit. Notice that, in the case of real tensors, the phase ambiguity reduces to a (positive/negative) sign ambiguity in the Schmidt basis vectors. A \emph{canonical form} for any acyclic network is defined by requiring that every internal index is in Schmidt form.

\begin{figure}[!t!b]
\begin{center}
\includegraphics[width=7cm]{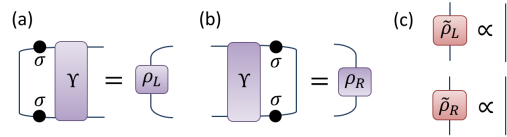}
\caption{(a-b) The left boundary matrix $\rho_L$ is formed from contracting a bond environment $\Upsilon$ with (two copies of) the associated bond matrix $\sigma$. (b) The right boundary matrix $\rho_R$. (c) The weighted trace gauge (WTG) is the choice of gauge that yields trivial environment matrices, $\tilde \rho_L \propto \mathbb I$ and $\tilde \rho_R \propto \mathbb I$. } 
\label{fig:CannonicalForm} 
\end{center}
\end{figure}

We now propose a means for fixing gauge degrees of freedom that is applicable to arbitrary internal indices, not only to bridges. In order to fix the gauge on an internal index of a tensor network $\T$, we first compute the corresponding bond environment $\Upsilon$ and bond matrix $\sigma$, as defined earlier. From these we define left and right boundary matrices, $\rho _L$ and $\rho _R$,
\begin{align}
  \left( {{\rho _L}} \right)_{i'}^i & = \sum\limits_{k,j,j'}  {\sigma _{kj}}{\sigma _{kj'}} \Upsilon _{j'i'}^{ji} \hfill \nonumber\\
  \left( {{\rho _R}} \right)_{j'}^j & = \sum\limits_{k,i,i'} {\Upsilon _{j'i'}^{ji}{\sigma _{ik}}} {\sigma _{i'k}}, \hfill \label{eq:1},
\end{align} 
see also Fig. \ref{fig:CannonicalForm}(a-b), which are symmetric and positive by construction (though not necessarily normalised with unit trace). Notice that, under change of gauge on the index under consideration, both corresponding the bond environment and bond matrix are themselves altered, with $\Upsilon \rightarrow \tilde \Upsilon$ and $\sigma \rightarrow \tilde \sigma$ as depicted in Fig. \ref{fig:GaugeChange}(f), which also changes $\rho _L$ and $\rho _R$. We now propose a particular choice of gauge: 
\begin{framed}
\textbf{Weighted trace gauge:} An index from a tensor network is in the weighted trace gauge (WTG) if associated the left and right boundary matrices $\rho_L$ and $\rho_R$ are proportionate to the identity, $\rho_L \propto \mathbb{I}$ and $\rho_R \propto \mathbb{I}$, and the bond matrix $\sigma$ is diagonal with positive elements in descending order, $\sigma_{ij} = s_i \delta_{ij}$ with $s_i \ge s_{i+1}$. The elements $s_i$ are henceforth referred to as the WTG coefficients of the index. We say the network is in \emph{canonical form} if all of the internal indices of the network are in the WTG.
\end{framed}
Before addressing a method to identify the gauge change matrices $x$ and $y$, see Fig. \ref{fig:GaugeChange}(d-e), that can bring an index into the WTG, we discuss some of its properties. Firstly, we note that if an internal index is a bridge of the network (or could realised as a bridge through a suitable unitary reorganisation of the external indices, see Sect. \ref{sect:A} of the appendix), then property \ref{E3} of the environment implies that the boundary matrix constraints of Eq. \ref{eq:1} are equivalent to left/right orthogonality condition of the Schmidt form. Thus for bridge indices, the WTG is precisely equivalent to the Schmidt gauge, such that the WTG coefficients are equal to the Schmidt coefficients. Notice also that, since the gauge condition on an index is only a property of the associated bond environment and bond matrix, properties \ref{E1} and \ref{E2} imply that the WTG on an index is invariant with respect to (i) the choice of gauge on other internal indices in the network and (ii) unitary transformation of the external indices. Implication (i) is particularly useful from an algorithmic standpoint, as it allows a network to be bought into canonical form by fixing the gauge on each internal index one at a time. 

\begin{figure}[!t!b]
\begin{center}
\includegraphics[width=8.5cm]{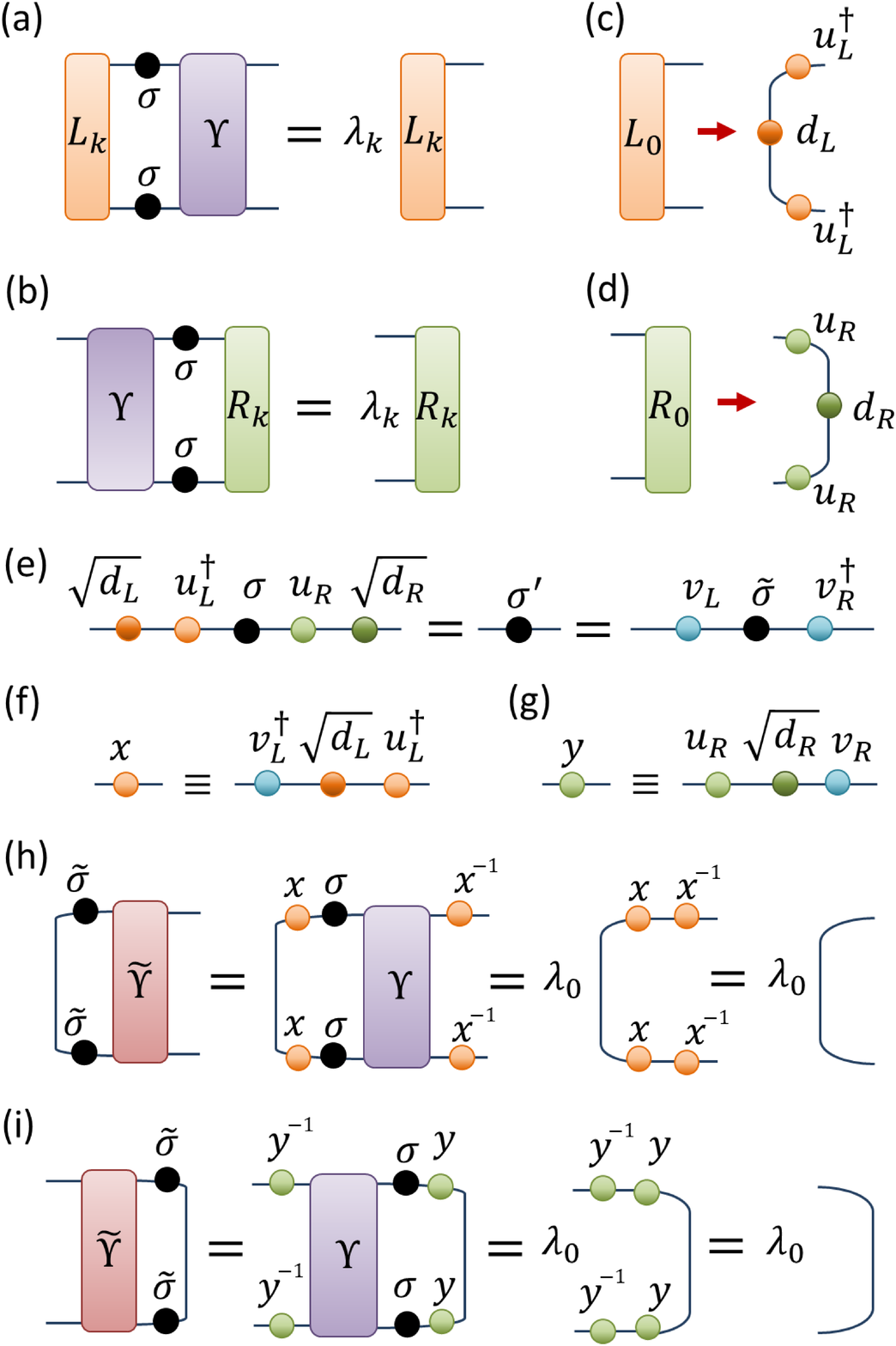}
\caption{Outline of the steps for fixing an internal index in the WTG, given the bond environment $\Upsilon$ and bond matrix $\sigma$. (a) Operator $L_k$ is a (left) eigenoperator of the transfer operator $(\sigma \otimes \sigma) \Upsilon$. (b) Operator $R_k$ is a (right) eigenoperator of the transfer operator $\Upsilon (\sigma \otimes \sigma)$. (c-d) Eigendecompositions of $L_0$ and $R_0$. (e) Definition of modified bond matrix $\sigma '$, which is then decomposed via the SVD, see also Eqs. \ref{eq:1e} and \ref{eq:1f}. (f-g) Definition of the gauge change matrices $x$ and $y$ that transform the index to the WTG. (h-i) Demonstration that the transformed environment $\tilde \Upsilon$ and bond matrix $\tilde \sigma$ satisfy the WTG constraints depicted Fig. \ref{fig:CannonicalForm}, i.e. that $\tilde \rho_L \propto \mathbb{I}$ and $\tilde \rho_R \propto \mathbb{I}$.} 
\label{fig:GaugeFixing} 
\end{center}
\end{figure}

We now propose a method of identifying the gauge change matrices $x$ and $y$ (and their inverses) that can bring an internal index into the WTG, given a bond environment $\Upsilon$ and bond matrix $\sigma$. We first form a (completely positive) transfer operator by contracting the bond weights to the left of the bond environment, $(\sigma \otimes \sigma) \Upsilon$, then diagonalize for the left dominant eigenoperator $L_0$,
\begin{equation}
L_0 (\sigma \otimes \sigma) \Upsilon = \lambda_0 L_0, \label{eq:1b}
\end{equation}
as depicted in Fig. \ref{fig:GaugeFixing}(a). Let us temporarily assume that the dominant eigenvalue is not degenerate, i.e. that $|\lambda_0| > |\lambda_k|$ for all other eigenvalues with $k>0$, where the degenerate case will be considered later. We similarly form a transfer operator by contracting the bond weights to the right of the bond environment, $\Upsilon (\sigma \otimes \sigma)$, then diagonalize for the right dominant eigenoperator $R_0$,
\begin{equation}
\Upsilon (\sigma \otimes \sigma) R_0 = \lambda_0 R_0, \label{eq:1c}
\end{equation}
see also Fig. \ref{fig:GaugeFixing}(b). 

Due to the symmetry of the transfer operators it follows that $L_0$ and $R_0$ are symmetric when viewed as matrices between their upper and lower indices, i.e. that $(L_0)^i_{i'} = (L_0)^{i'}_i$ and $(R_0)^i_{i'} = (R_0)^{i'}_i$, such that they can be diagonalized
\begin{align}
{L_0} &= {u_L} {{d_L}} u_L^\dag, \nonumber \\
{R_0} &= {u_R} {{d_R}} u_R^\dag, \label{eq:1d}
\end{align}
see Fig. \ref{fig:GaugeFixing}(c-d), with unitary matrices $u_L$, $u_R$ and real diagonal matrices $d_L$, $d_R$. Notice that, due to the positivity of the bond environment $\Upsilon$, it follows from the Perron-Frobenius theorem that $d_L$ and $d_R$ are non-negative, thus possess real roots $\sqrt {{d_R}}$ and $\sqrt {{d_R}}$. We now use these to transform the bond matrix $\sigma$, 
\begin{equation}
\sigma ' \equiv \sqrt {{d_L}} u_L^\dag \sigma \  u_R \sqrt{{d_R}}, \label{eq:1e}
\end{equation}
and take the singular value decomposition to obtain
\begin{equation}
\sigma ' = {w_L}\tilde \sigma w_R^\dag \label{eq:1f}
\end{equation}
for unitary $w_L$, $w_R$ and positive diagonal $\tilde \sigma$. The gauge change matrices $x$ and  $y$ that bring the index into the WTG can now be defined as
\begin{align}
x &\equiv w_L^\dag \sqrt {{d_L}} u_L^\dag, \nonumber\\
y &\equiv u_R \sqrt {{d_R}} w_R, \label{eq:1g}
\end{align}
see also Fig. \ref{fig:GaugeFixing}(f-g), where one should notice that $L_0 = x^\dag x$ and $R_0 = y y^\dag$. Under this choice of gauge the new bond matrix is simply the $\tilde \sigma$ from Eq. \ref{eq:1f}, which is positive and diagonal by construction. Furthermore the new bond environment $\tilde \Upsilon$, obtained after the gauge transformation on $\Upsilon$ as shown in Fig. \ref{fig:GaugeChange}(d-f), is seen to satisfy the remaining WTG constraints, such that the left and right boundary matrices are proportionate to identity, as illustrated in Fig. \ref{fig:GaugeFixing}(h-i).  

We now discuss the existence and uniqueness of the WTG. First we note that, while the matrices $x$ and $y$ of Eq. \ref{eq:1g} are always well-defined, they can only constitute a valid change of gauge if they are invertible, which implies that the WTG exists if and only if the matrices ${L_0}$ and ${R_0}$ of Eq. \ref{eq:1d} have strictly positive eigenvalues (whereas the Perron-Frobenius theorem only guarantees that they are non-negative). In regards to uniqueness one should note that the WTG is not unique if there exists degeneracy between two or more of the WTG coefficients in $\tilde \sigma$ as unitary gauge freedom within degenerate subspace still remains (identical to the gauge freedom in the Schmidt form when there exists degeneracy in the Schmidt coefficients). Similarly the WTG also has the same phase ambiguity as discussed earlier in the context of the Schmidt decomposition. Notice that both of these freedoms relate potential ambiguities in the singular value decomposition of Eq. \ref{eq:1f}. However, aside from these, the WTG is unique provided that the dominant eigenvalues of the transfer operators in Eq. \ref{eq:1b} and Eq. \ref{eq:1c} are non-degenerate; this follows as (i) the dominant eigenoperators $L_0$ and $R_0$ are the only eigenoperators that can have a strictly positive eigendecomposition in Eq. \ref{eq:1d} and (ii) any ambiguity in this eigendecomposition can be absorbed into the later singular value decomposition step. However, if the transfer operators of Fig. \ref{fig:GaugeFixing}(a-b) do have degeneracy in the dominant eigenvalue, i.e. such that $\lambda_0 = \lambda_1$, then one could use any linear combination of the dominant eigenoperators provided that they possess a strictly positive decomposition in Eq. \ref{eq:1d}; thus the WTG would not be unique.

Finally, it is worth remarking on the similarity between gauge-fixing procedure described above and methods for gauge-fixing in matrix product states (MPS)\cite{TN2, TN3}. Let us consider a translation invariant matrix product state for an infinite $1D$ lattice, composed of translations of identical three index tensors $A$. It can be seen that the method for finding the gauge that brings an internal index to the WTG is precisely equivalent to the standard approach for bringing the MPS into Schmidt form if one substitutes the bond environment $\Upsilon$ with the standard transfer matrix $T$ of the MPS formed from contracting tensor $A$ with its conjugate. Thus the conditions of existence and uniqueness for WTG are also precisely the same as the conditions for the existence and uniqueness of the Schmidt gauge (or canonical form) of an infinite MPS.

\begin{figure}[!t!b]
\begin{center}
\includegraphics[width=8.5cm]{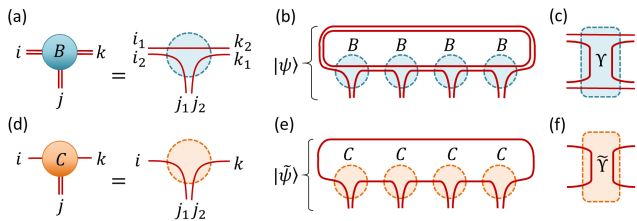}
\caption{(a) Tensor $B_{ijk}$ and its internal structure, see also Eq. \ref{eq:B}. (b) State $\ket{\psi}$ given by a periodic MPS composed of four copies of $B$. (c) Bond environment from an internal index in $\ket{\psi}$. (d) Tensor $C_{ijk}$ and its internal structure, see also Eq. \ref{eq:C}. (e) State $\ket{\tilde \psi}$ given by a periodic MPS composed of four copies of $C$. (f) Bond environment from an internal index in $\ket{\tilde \psi}$.} 
\label{fig:CDL}
\end{center}
\end{figure}

\section{Internal correlations} Given that the WTG simplifies to the Schmidt form in acyclic networks, one may be tempted to believe that the WTG coefficients of an index directly relate to some physical property of the quantum state described by a tensor network, just as the Schmidt coefficients relate to the bipartite entanglement entropy of a quantum state\cite{Neilson}. However it turns out that this is not the case, due to the possibility of \emph{internal correlations} within tensor networks that contains closed loops, as we now demonstrate with a simple example.  

Let $B_{ijk}$ be a three index tensor where each index is of dimension $\chi=4$, such that $i,j,k \in \{1,2,3,4 \}$. It follows that each index can be decomposed a product of two finer indices of dimension $2$, i.e. $i = 2*i_2 + i_1$ with $i_1, i_2 \in \{1,2 \}$. Tensor $B$ is then defined as having $\delta$-function correlations on the finer indices,
\begin{equation}
B_{({i_1}{i_2})({j_1}{j_2})({k_1}{k_2})} = {\delta _{{i_2}{j_1}}}{\delta _{{j_2}{k_1}}}{\delta _{{k_2}{i_1}}} \label{eq:B}
\end{equation}
as also depicted in Fig. \ref{fig:CDL}(a). Let us now construct a periodic MPS, denoted as $\T$, of bond dimension $\chi=4$ formed from four copies of $B$, as depicted in Fig. \ref{fig:CDL}(b). Notice that the network $\T$ represents a quantum state $\ket{\psi}$ that consists of a product of nearest neighbour singlets. Similarly we define a new tensor $C_{i_1 jk_1}$, 
\begin{equation}
C_{{i_1}({j_1}{j_2}){k_1}} = {\delta _{{i_1}{j_1}}}{\delta _{{j_2}{k_1}}} \label{eq:C}
\end{equation}
where the indices are of dimension $\chi=2$, i.e. $i_1, j_1, j_2, k_1 \in \{1,2 \}$, with the index $j = 2*j_2 + j_1$. We then form a periodic MPS, denoted $\tilde \T$, from four copies of $C$ as depicted in Fig. \ref{fig:CDL}(c). It is easily understood that the quantum state $\ket{\tilde \psi}$ described by the network $\tilde \T$ is again a product of nearest neighbour singlets, identical to the previous state $\ket{\psi}$ up to normalization. However, despite describing the same quantum state, the two tensor network representations $\T$ and $\tilde \T$ are fundamentally different; network $\T$ contains a string of \emph{internal correlations} around the closed loop, although these do not contribute to any property of the corresponding quantum state. It is also easily checked that the WTG coefficients differ between the two tensor network representations; network $\T$ has four equal WTG coefficients on each index versus two equal coefficients in $\tilde \T$. This is a clear demonstration that the WTG coefficients of a cyclic tensor network do not necessarily correspond to a physical property of the quantum state represented by a tensor network. One can understand this as a consequence of the inability of the WTG coefficients to distinguish between (physically meaningful) correlations between external indices and (physically irrelevant) internal correlations around closed loops.
 
The possibility of such internal correlations marks an important distinction between acyclic and cyclic tensor networks. We now propose a way to quantify the presence of internal correlations through an internal index of a tensor network. The are some natural criteria that a such measure should satisfy: (i) it should be zero if index under consideration is a bridge and (ii) it should be invariant under choice of gauge on the index. In order to arrive at such a measure, we compute the eigenvalues $\lambda_\alpha$ of the transfer operator formed by contracting the bond weights to the left of the bond environment, $(\sigma \otimes \sigma) \Upsilon$, as depicted in Fig. \ref{fig:GaugeFixing}(a). 
Notice the eigenvalues $\lambda_\alpha$ are clearly invariant under a change of gauge on the index under consideration (as this is equivalent to a change of basis on the transfer operator). We now take the absolute value of the eigenvalues (which, in general, may be complex) and normalise them, $\tilde \lambda_\alpha \equiv |\lambda_\alpha | / \left( {\sum\nolimits_\alpha |{\lambda_\alpha}|} \right)$, and define the cycle entropy $S_\textrm{cycle}$ as the von-Neumann entropy of this normalized spectrum, 
\begin{equation}
S_\textrm{cycle} = -\sum\nolimits_{\alpha} \left( {\tilde \lambda}_\alpha \log_2 \left( {\tilde \lambda}_\alpha \right) \right).  
\end{equation}
Notice that, by property \ref{E3} of the bond environment, it is clear that $S_\textrm{cycle} = 0$ if the index under consideration is a bridge, as desired. The reverse statement is also true: if $S_\textrm{cycle} = 0$ then it follows that the index under consideration can be realised as a bridge, perhaps after some appropriate unitary cycle reduction as discussed in Sect. \ref{sect:A} of the appendix. Thus, if the cycle entropy is zero, then the WTG coefficients are precisely equal to the Schmidt coefficients of this bridge realization. It follows that if the cycle entropy $S_\textrm{cycle}$ through an index is zero (or sufficiently small) one can achieve an optimal (or near optimal) truncation of this index by transforming to the WTG and then simply discarding the smallest WTG coefficients. This demonstrates a useful application of the cycle entropy $S_\textrm{cycle}$ in quantifying the extent of internal correlations through an index of a tensor network. 

\section{Optimal truncations} A task that ubiquitous in tensor network algorithms is that of truncating an internal index from some initial dimension $\chi$ to some smaller dimension $\tilde \chi < \chi$ in an optimal manner. In the case of bridge indices this is easily accomplished be discarding the smallest of its corresponding Schmidt coefficients. A common approach to the truncation of non-bridge indices is to reduce the index under consideration to a bridge by ``cutting" open other indices of the network, a process we refer to as a cycle reduction via cutting in Sect. \ref{sect:B} of the appendix, and then applying a Schmidt decomposition. However if the cycle entropy $S_\textrm{cycle}$ through an internal index is non-zero, then such a cycle reduction will not produce an optimal truncation. In this case a more sophisticated approach is required, one which can distinguish and remove the redundant internal correlations from the network. We now propose an algorithm that can potentially achieve an optimal truncation even for internal indices which have non-zero cycle entropy, $S_\textrm{cycle} \ne 0$, which we call a \emph{full environment truncation} (FET). 

\begin{figure}[!t!b]
\begin{center}
\includegraphics[width=8cm]{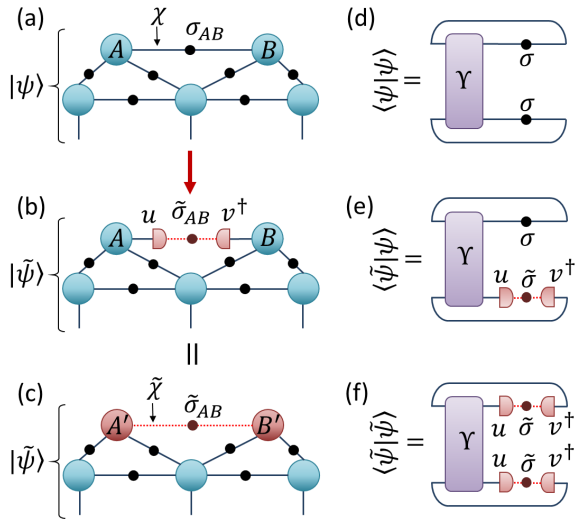}
\caption{(a-c) The index connecting tensors $A$ and $B$ is truncated to smaller dimension, $\tilde \chi < \chi$, by replacing $\sigma_{AB}$ with the product $u ({\tilde \sigma}_{AB}) v^\dag$, where $u,v$ are isometries and ${\tilde \sigma}_{AB}$ is a $\tilde \chi \textrm{-by-} \tilde \chi$ matrix. (d-f) The overlaps $\braket{\psi}{\psi}$, $\braket{\tilde \psi}{\psi}$, $\braket{\tilde \psi}{\tilde \psi}$ of the states from (a-b), which have been expressed in terms of the bond environment $\Upsilon$.}
\label{fig:Truncate}
\end{center}
\end{figure}

Let us assume that we have a tensor network $\T$, describing a quantum state $\ket{\psi}$, and that we wish to optimally truncate the dimension of a chosen internal index from initial dimension $\chi$ to final dimension $\tilde \chi < \chi$, i.e. as to leave the resulting state $\ket{\phi}$ as close to the original as possible. Here we quantify the difference between the initial state $\ket{\psi}$ and the final state $\ket{\phi}$ of the truncated network using the fidelity $F$,
\begin{equation}
F\left( \psi ,{ \phi} \right) = \frac{\braket{ \phi}{\psi} \braket{\psi}{\phi}}{\braket{ \phi}{ \phi} \braket{\psi}{\psi}}, \label{eq:F}
\end{equation}
which we seek to maximise. The truncation can be implemented by replacing the bond matrix $\sigma$ of the index under consideration with some rank $\tilde \chi$ matrix which, making use of the SVD, can generically be expressed as the product $u{\tilde \sigma} v^\dag$, see Fig. \ref{fig:Truncate}(b). Here $u$ and $v$ are $\chi \textrm{-by-} \tilde \chi$ isometries, such that $u^\dag u = v^\dag v = \mathbb{I}_\chi$, with $\mathbb{I}_\chi$ the $\chi \textrm{-by-} \chi$ identity matrix, and $\tilde \sigma$ is a $\tilde \chi \textrm{-by-} \tilde \chi$ diagonal matrix of positive real values. Notice that the isometries $u$ and $v$ can be absorbed into their adjoining tensors respectively, see Fig. \ref{fig:Truncate}(c), such that the resulting tensor network $\tilde \T$, which defines the new quantum state $\ket{\phi}$ is of the same geometry as the original but with a reduced index dimension of $\tilde \chi$. The task of optimally truncating an internal index can thus be recast as optimizing isometries $u$, $v$ and a matrix $\tilde \sigma$ such as to maximise the fidelity of Eq. \ref{eq:F}. We now describe an outline of an iterative optimization algorithm for isometries $u$, $v$ and the matrix $\tilde \sigma$, the full details of which can be found in Sect. \ref{sect:D} of the appendix. For the first step we define $R \equiv ( {\tilde \sigma} v^\dag)$, then solve for the optimal $R$ while the $u$ tensor is held fixed. This can be achieved through standard techniques, as it is equivalent to solving a generalized eigenvalue problem for $R$. Once the optimal $R$ is obtained the SVD is taken to produce updated tensors $\tilde \sigma$ and $v$. At the next step, the product $L \equiv (u {\tilde \sigma})$ is similarly updated with $v$ held fixed. These two steps are iterated until the all tensors converge. Notice that the terms in the fidelity of Eq. \ref{eq:F} can be expressed solely using the corresponding bond environment $\Upsilon$ and bond matrix $\sigma$, as depicted in Fig. \ref{fig:Truncate}(d-f). Thus the FET optimization algorithm only requires these two tensors as an input, and can be applied regardless of the wider structure of the network under consideration (so long as the environment $\Upsilon$ can be computed).

\begin{figure}[!t!b]
\begin{center}
\includegraphics[width=8cm]{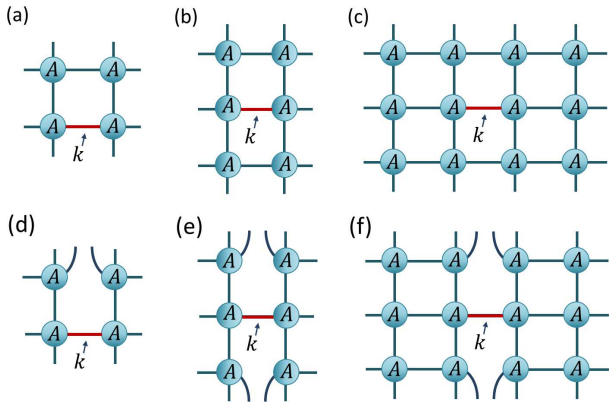}
\caption{(a) $2\textrm{-by-}2$, (b) $3\textrm{-by-}2$ and (c) $3\textrm{-by-}4$ networks of tensors $A$. (d-f) The index $k$ can be turned into a bridge index by appropriately cutting other network indices.}
\label{fig:IntCutter}
\end{center}
\end{figure}

In order to test the FET algorithm we apply it to the partition function of the (classical) square-lattice Ising model at critical temperature $T_c$. We begin from the standard tensor network representation of the partition function, where each four index tensor $A_0$ represents the Boltzmann weights of a plaquette of Ising spins\cite{TNR4}. Then we form coarse-grained tensors $A$ through application of four iterations of the higher order tensor renormalization group (HOTRG) algorithm\cite{TRG6} (with bond dimension limited at $\chi = 16$), such that each $A$ now represents the Boltzmann weights of a (coarse-grained) $16\textrm{-by-}16$ block of Ising spins. Finally, we apply the closed-loop truncation algorithm to truncate an internal index of $2\textrm{-by-}2$, $3\textrm{-by-}2$ and $3\textrm{-by-}4$ blocks of $A$ tensors, as depicted in Fig. \ref{fig:IntCutter}(a-c), from initial dimension $\chi = 16$ to final dimension $\tilde \chi = 4$. The results of this test for the error in the fidelity, $\epsilon = 1-F$, are displayed in Tab. \ref{tab:Finite}. We compare between the error $\epsilon_\textrm{\tiny CR}$ from a Schmidt decomposition applied to a cycle reduction (see Fig. \ref{fig:IntCutter}(d-f)), to the error $\epsilon_\textrm{\tiny FET}$ from the FET algorithm. In all instances it is seen that the FET is more accurate, with $\epsilon_\textrm{\tiny FET} < \epsilon_\textrm{\tiny CR}$, and that the magnitude of the accuracy improvement grows as the cycle entropy $S_\textrm{cycle}$ increases. This is as expected, since the FET algorithm achieves a more accurate truncation through removal of internal correlations, seen in the reduction of the cycle entropy $S_\textrm{cycle}$ in Tab. \ref{tab:Finite}, whereas the cycle reduction approach preserves the internal correlations. In all cases, less than 20 iterations were required to optimise the tensors necessary for the FET algorithm. It was also found that the optimizations seemed to converge to the same final tensors regardless of how they were first initialized. This  seems to suggest that the FET algorithm is converging to the global minimum in the fidelity error, as opposed to getting stuck in a local minima.  

\begin{table}[!t]
\begin{tabular}{|l||c|c|c|}
  \hline
                           & $\epsilon_\textrm{\tiny CR}$ & $\epsilon_\textrm{\tiny FET}$ & $S_\textrm{cycle}$ \\ \hline
   (a) $\ 2\times 2 \ $ &  $\ 6.7\times 10^{-4} \ $     &  $\ 5.0\times 10^{-5} \ $     & $\; 1.37 \rightarrow 1.04 \; $ \\ 

   (b) $\ 3\times 2 \ $ &  $\ 2.0\times 10^{-5} \ $     &  $\ 1.0\times 10^{-8} \ $     & $\; 2.27 \rightarrow 2.15 \; $ \\

   (c) $\ 3\times 4 \ $ &  $\ 7.2\times 10^{-6} \ $     &  $\ 5.0\times 10^{-10} \ $     & $\; 2.31 \rightarrow 2.22 \; $ \\

  \hline
  \end{tabular} 
  \caption{Fidelity errors from truncation of index $k$ from initial dimension $\chi = 16$ to final dimension $\tilde \chi = 4$ in the networks of Fig. \ref{fig:IntCutter}(a-c), comparing error $\epsilon_\textrm{\tiny FET}$ from a full environment truncation (FET) to the error $\epsilon_\textrm{\tiny CR}$ from a Schmidt decomposition of the cycle reductions depicted Fig. \ref{fig:IntCutter}(d-f). The cycle entropy $S_\textrm{cycle}$ of index $k$ before and after the FET is also given.}
  \label{tab:Finite}
\end{table}

\section{Discussion}
While tensor networks that contain closed loops are undoubtedly more complicated than their loop-free counterparts, this manuscript has introduced several ideas to facilitate working with such networks. These include: (i) a method of fixing the gauge degrees of freedom (which yields a well-defined canonical form for arbitrary networks), (ii) a means of quantifying the extent of internal correlations through an index (via the cycle entropy $S_\textrm{cycle}$), and (iii) an algorithm that potentially allows for the optimal truncation of indices through removal of internal correlations. We envision these results will have useful applications across a broad range of tensor network algorithms, some of which we discuss below.

Fixing the gauge is particularly useful in optimisation algorithms for tensor networks as it can allow certain intermediate tensors to be reused between optimisation iterations. For instance, a large amount of computation time in the iPEPS algorithm\cite{PEPS1,PEPS2,PEPS3} is spent computing the boundary MPS, necessary for evaluation of the local environment, which must be recomputed every time the PEPS tensors change. Fixing the gauge allows the boundary MPS from a previous iteration to be used as the starting point for the calculation of the updated boundary MPS (whereas, if the gauge were not properly fixed, then the previous boundary MPS may be in a different gauge and thus not suitable as the starting point). For the specific case of translation invariant iPEPS an alternative means of fixing the gauge was already proposed in Ref.\onlinecite{PEPS4}, which allowed for significant improvements to the efficiency through proper recycling of the environment. The results of this manuscript provide a more general way to accomplish this task of recycling intermediate tensors, which could be applied to arbitrary networks.

The measure $S_\textrm{cycle}$ for quantifying extent of internal correlations, and the FET scheme for removing internal correlations, are directly applicable to tensor renormalization group (TRG) schemes for coarse-graining path integrals and partition functions. A significant problem with the original TRG scheme of Levin\cite{TRG1}, and its later generalizations\cite{TRG2,TRG3,TRG4,TRG5,TRG6,TRG7}, is that they fail to remove internal correlations. These internal correlations can thus accumulate over successive RG steps and cause a computational break down of the approach. This problem was resolved with tensor network renormalization\cite{TNR1,TNR2,TNR3,TNR4} (TNR), which introduced unitary disentanglers to remove internal correlations and prevent their accumulation, allowing a sustainable RG flow. Many similar methods have followed\cite{TNRc1,TNRc2,TNRc3,TNRc4,TNRc5}, using a variety of alternate techniques to remove internal correlations as part of the coarse-graining step. Likewise the FET algorithm, which was demonstrated to be effective in the removal of internal correlations, can directly be incorporated as part of a TNR-like renormalization scheme for tensor networks. The details of this implementation and some benchmark results are described in Sect. \ref{sect:E} of the appendix. For the $2D$ classical Ising model at critical temperature, this approach was able to resolve the free energy per spin with an accuracy of $\delta f < 3 \times 10^{-10}$ on a lattice of $2^{16}\times 2^{16}$ spins. This calculation required approximately 20 minutes computation time on a desktop PC, which compares favourably with previous approaches. A key feature of the FET is that it can be applied to remove internal correlations from arbitrary networks regardless of their geometry, similar to the recently proposed approach of Ref. \onlinecite{TNRc4}, in contrast to most other previous approaches which are specialised to a single geometry. For instance, the proposed approach could straight-forwardly be generalized to coarse-grain $3D$ networks, allowing $2D$ quantum systems to be studied, which will be considered in future work.   

The author thanks David Poulin and Markus Hauru for useful discussions. This research was undertaken thanks in part to funding from the Canada First Research Excellence Fund.

\appendix
\newpage

\section{Cycle reductions via external unitaries} \label{sect:A}
In this Appendix we discuss examples of tensor networks networks where a non-bridge internal index (i.e. an index that is contained in a cycle) can be reduced to a bridge via a suitable external unitary transformation $U$, which we call a unitary cycle reduction.

\begin{figure}[htb]
\begin{center}
\includegraphics[width=8.0cm]{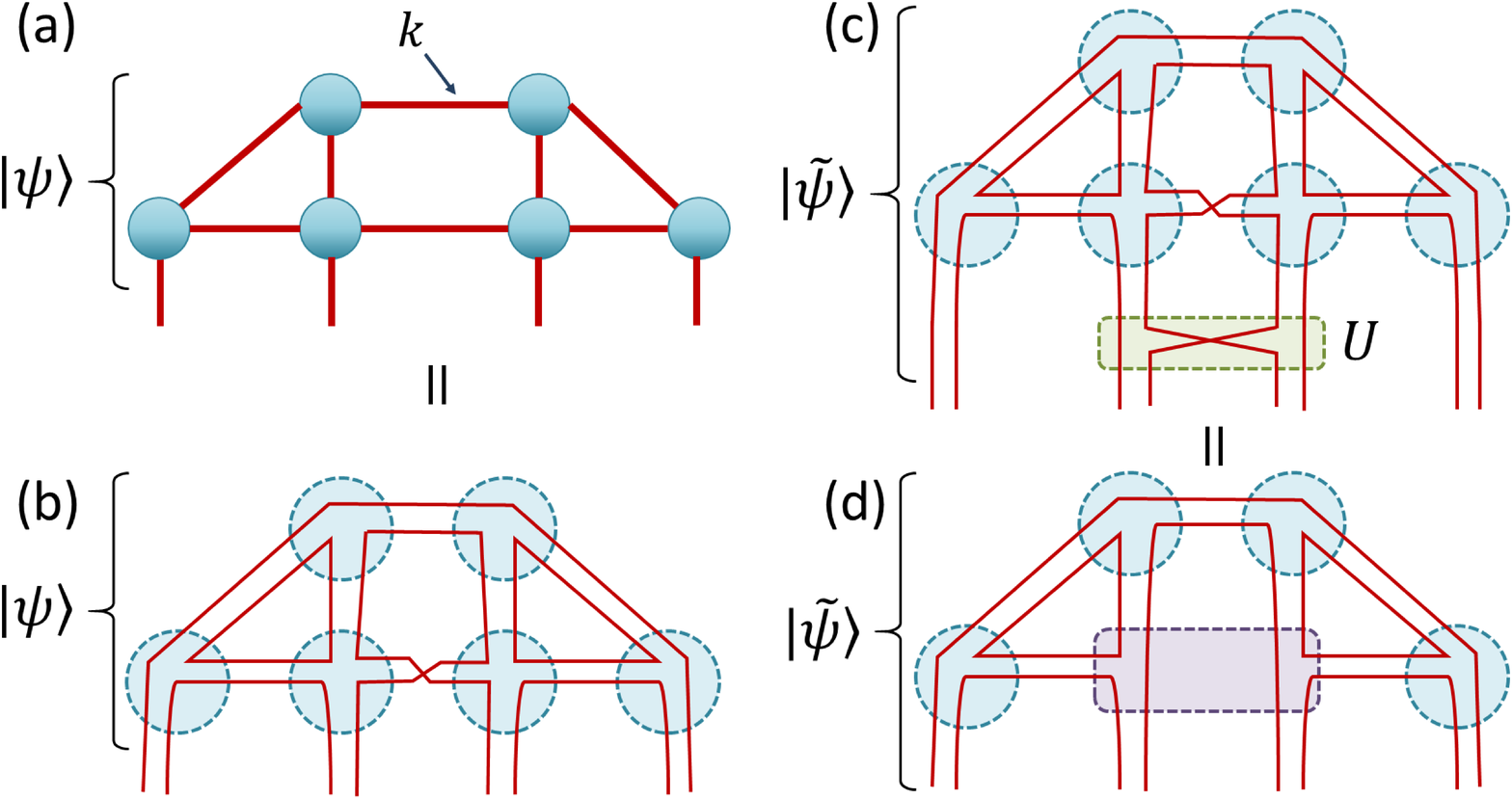}
\caption{(a) A tensor network describes a quantum state $\ket{\psi}$ on a lattice of four sites. (b) A specific instance of the network from (a), where each tensor index can be decomposed as a product of two finer indices, and that the tensors have $\delta$-function like correlations in the finer indices as depicted. (c-d) Application of a suitably chosen unitary $U$ on the external indices allows index $k$ to be realised as a bridge of the network.} 
\label{fig:UnitaryCut}
\end{center}
\end{figure}

\begin{figure}[htb]
\begin{center}
\includegraphics[width=8.0cm]{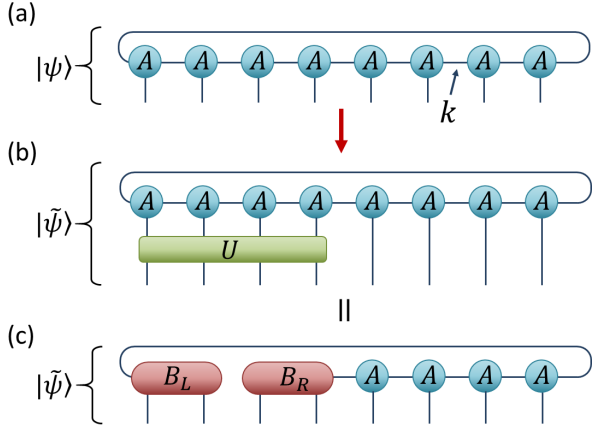}
\caption{(a) A periodic MPS, which describes a quantum state $\ket{\psi}$. We assume that the MPS is injective and has a small correlation length, $\zeta \ll 4$. (b-c) An appropriately chosen unitary $U$ may disentangle the MPS to an acyclic network.} 
\label{fig:ExtUnitary}
\end{center}
\end{figure}

Consider the example presented in Fig. \ref{fig:UnitaryCut}(a-b); here a network, composed of corner double line (CDL) tensors, describes a quantum state $\ket{\psi}$ on a four site lattice. It is easily seen that, for the labelled index $k$, there is no possible partitioning of the external indices such that $k$ is a bipartition of the state $\ket{\psi}$. However, in this example, there exists a unitary $U$ such that in the transformed state, $\ket{\tilde \psi} = U \ket{\psi}$, the index $k$ has become a bridge, as depicted in Fig. \ref{fig:UnitaryCut}(c-d). A second example, consisting of a periodic MPS, is presented in Fig. \ref{fig:ExtUnitary}. We assume that the MPS is injective and has a finite correlation length $\zeta$, as would be the case if the MPS described the ground state of a gapped periodic system. Then one may argue that there exists some unitary $U$, acting on $L > \zeta$ sizes, that would reduce the periodic MPS to an acyclic network as depicted in Fig. \ref{fig:ExtUnitary}(b-c). 

Once an internal index has been reduced to a bridge then its gauge may be fixed or its dimension truncated using the Schmidt decomposition. The results from the main text are particular useful in characterising when it is possible for an internal index to become a bridge; there exists an external unitary $U$ that allows an internal index to become a bridge if and only if the corresponding cycle entropy is zero, $S_\textrm{cycle} = 0$. However, in instances that $S_\textrm{cycle} = 0$, the Schmidt gauge that would be reached after the external unitary $U$ is precisely equivalent to the WTG, but the latter can be determined without first needing to determine $U$. 
 
\section{Cycle reductions via index cutting}
\label{sect:B}
\begin{figure}[tb]
\begin{center}
\includegraphics[width=8.5cm]{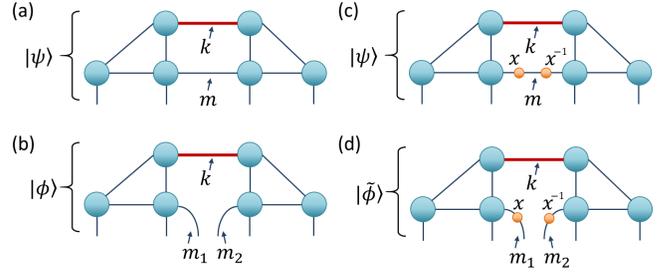}
\caption{(a) A tensor network describes a quantum state $\ket{\psi}$. (b) Internal index $m$ is cut as to become a pair of external indices $m_1$ and $m_2$, such that internal index $k$ becomes a bipartition of the resulting network. (c) A gauge transformation is made on index $m$ from the network of (a). (d) The state $\ket{\tilde \phi}$ produced from cutting index $m$ after the gauge transformation differs from the state $\ket{\phi}$ from (b).} 
\label{fig:IndexCut}
\end{center}
\end{figure}

Discussed in this appendix is a standard way of dealing with gauge-fixing and truncation of non-bridge indices in cyclic tensor networks, whereby the index under consideration is made into a bridge by ``cutting" indices of the original network. An example of this is given in Fig. \ref{fig:IndexCut}, where it is assumed that one wants to fix the gauge and/or truncate index $k$ from a network describing quantum state $\ket{\psi}$. This can be achieved by cutting internal index $m$, thus promoting it to a pair of external indices $m_1$ and $m_2$, see Fig. \ref{fig:IndexCut}(b). Notice that index $k$ is now a bridge of the new tensor network, which now describes a quantum state $\ket{\phi}$ in an enlarged Hilbert space. We call this manipulation a cycle reduction (via cutting) of the network with respect to $k$. One could then fix the gauge on index $k$ and truncate its dimension using a Schmidt decomposition on the reduced network. 

However, there are several significant problems with the cycle reductions based on cutting. The first problem is that they are not unique. Consider, for instance, the example given in Fig. \ref{fig:IndexCut}(c) where a change of gauge on the internal index $m$ changes the quantum state produced by the reduction, see Fig. \ref{fig:IndexCut}(d), thus also changes the Schmidt basis on index $k$. More generally, one also has freedom in the choice of which internal indices are cut to produce the reduction. The second, perhaps more severe, problem is that this type of cycle reductions does not (in general) allow for an optimal truncation of an internal index. This follows as the cycle reduction will promote internal correlations into physical correlations (in the enlarged Hilbert space), such that they will be preserved in the subsequent Schmidt decomposition. In contrast, a method which takes the internal correlations in account, such as the FET algorithm presented in the main text, can potentially achieve a more accurate truncation through identification and removal of internal correlations. This can be seen in Tab. \ref{tab:Finite}, which compares truncation of the networks depicted Fig. \ref{fig:IntCutter} based on FET against truncation based on cycle reductions.

\section{Algorithm for optimal internal truncations}
\label{sect:D}
\begin{figure}[!t!hb]
\begin{center}
\includegraphics[width=8.0cm]{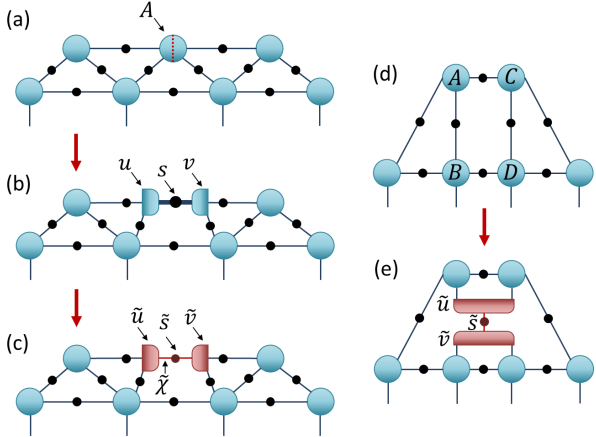}
\caption{(a-c) Tensor $A$ is first decomposed into a product of tensors via the SVD, and then the connecting index is truncated down to a smaller dimension $\tilde \chi$. (d-e) A pair of indices in the network is truncated to a single effective index.} 
\label{fig:ManyTruncate}
\end{center}
\end{figure}

In this appendix we describe the algorithm for a full environment truncation (FET), which allows for a potentially optimal truncation of an internal index in a network. Before introducing this algorithm we note that, although formulated as a method for truncating a single internal index, the FET method can be easily applied to several alternate scenarios. For instance, the FET could also be applied if one wanted to split a single tensor $A$ from a network into a pair of tensors in an optimal manner. Here one can first decompose the tensor $A$ using a (truncation-free) singular value decomposition, and then apply the FET algorithm to truncate the connecting index, as depicted in Fig. \ref{fig:ManyTruncate}(a-c). One can also apply the  FET algorithm to truncate multiple internal indices down to a single effective index, as depicted in Fig. \ref{fig:ManyTruncate}(d-e).

\begin{figure}[!thb]
\begin{center}
\includegraphics[width=8.5cm]{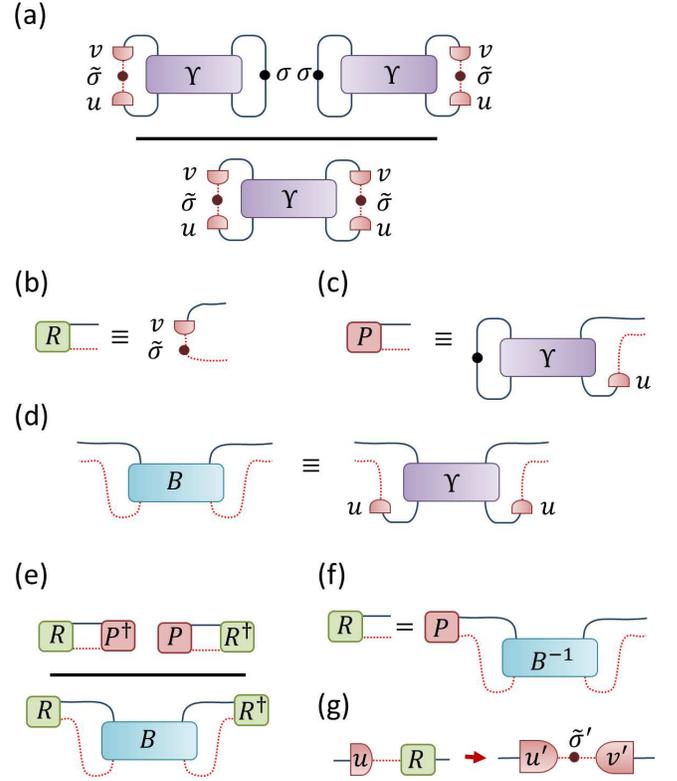}
\caption{Diagrams relating to the full environment truncation (FET) algorithm. (a) The fidelity $F$ between an initial and a truncated state expressed in terms of a bond environment $\Upsilon$, see Fig. \ref{fig:Truncate}. (b-d) Definitions of tensors $R$, $P$ and $B$. (e) The fidelity $F$ can be expressed as a generalized eigenvalue problem in $R$ as $F = \left(R P^\dag P  R^\dag \right) / \left( R  B R^\dag \right)$. (f) The fidelity is maximized with the choice $R = P B^{-1}$. (g) Updated tensor $u'$, $\tilde \sigma'$ and $v'$ are obtained from the SVD of the product $u R$.} 
\label{fig:TruncateAlg}
\end{center}
\end{figure}

As discussed in the main text, the problem of truncating an internal index $k$ of a tensor network from some initial dimension $\chi$ to a smaller dimension $\tilde \chi$ can be reformulated as one of replacing the bond matrix $\sigma$ on $k$ with a product of tensors $u{\tilde \sigma} v^\dag$ as depicted in Fig. \ref{fig:Truncate}(b). Here $u$ and $v$ are $\chi \textrm{-by-} \tilde \chi$ isometries, such that $u^\dag u = v^\dag v = \mathbb{I}_\chi$, with $\mathbb{I}_\chi$ the $\chi \textrm{-by-} \chi$ identity matrix, and $\tilde \sigma$ is a diagonal matrix of positive real values. We now propose an iterative algorithm to find these tensors to maximise the fidelity, see Eq. \ref{eq:F}, of the truncated state with the original state. Before starting the iterations, we compute the bond environment $\Upsilon$ of the index under consideration, which allows the fidelity $F$ to be expressed as a simple quotient of tensor networks containing $\Upsilon$, see Fig. \ref{fig:TruncateAlg}(a). One should then initialise the tensors $\{ u, {\tilde \sigma}, v^\dag\}$, which can be done in a number of ways. Perhaps the simplest initialization is achieved by taking a truncated SVD of the bond matrix $\sigma$, retaining the $\tilde \chi$ largest singular values. In the first step we define $R \equiv ( {\tilde \sigma} v^\dag)$, and then seek to solve for the optimal $R$ while the $u$ tensor is held fixed. Let us define tensors $P$ and $B$ from the environment $\Upsilon$ as depicted in Fig. \ref{fig:TruncateAlg}(c-d). This allows us to write express the fidelity as a generalized eigenvalue problem in $R$,
\begin{equation} 
F = \left(R A  R^\dag \right) / \left( R  B R^\dag \right) \label{eq:D1}
\end{equation} 
with $A = P^\dag P$, see also Fig. \ref{fig:TruncateAlg}(e). Given that $A$ is simply the outer product of vectors $P$, the solution for $R$ that maximises the fidelity of Eq. \ref{eq:D1} is known analytically as $R_\textrm{max} = P B^{-1}$. One can then take the SVD of $R_\textrm{max}$ to obtain updated $\tilde \sigma$ and $v$ tensors, see Fig. \ref{fig:TruncateAlg}. At the next step, the product $L \equiv (u {\tilde \sigma})$ is similarly updated with $v$ held fixed. These two steps should be iterated until the tensors converge sufficiently. In the examples considered in the main text convergence required less than 20 iterations. 

\begin{figure}[htb]
\begin{center}
\includegraphics[width=8.0cm]{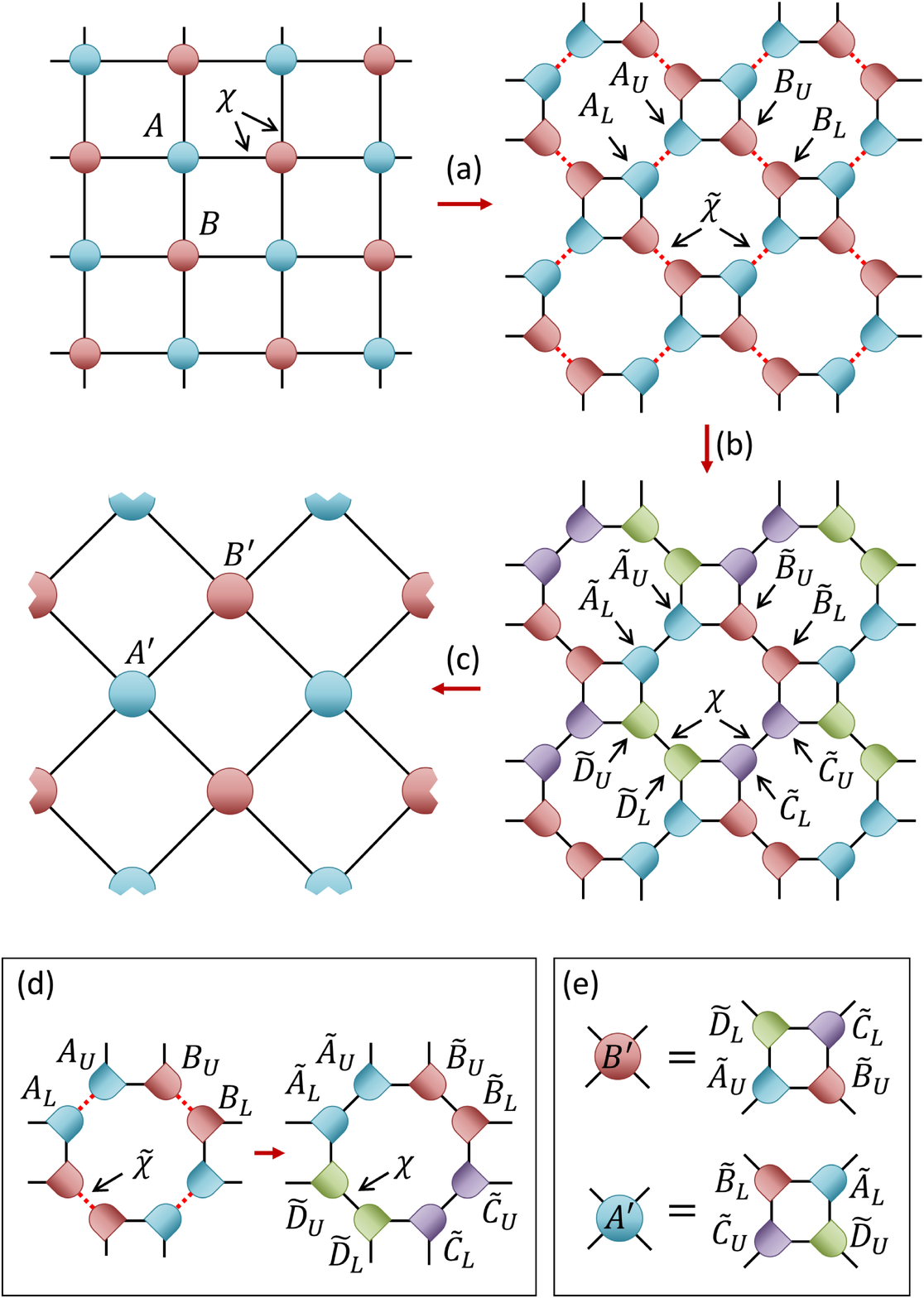}
\caption{An iteration of a coarse-graining algorithm for a square lattice network, which uses the FET approach to reduce internal correlations. (a) Tensors $A$ and $B$ are decomposed into products of 3-index tensors using the singular value decomposition, where $\tilde{\chi}$ singular values have been retained. (b) The closed-loop truncation scheme is applied to a section of the network containing a loop of 8 tensors, in order to truncate indices of dimension $\tilde{\chi}$ to smaller dimension $\chi < \tilde{\chi}$, as illustrated in (d). (c) A coarser square-lattice network is formed through the contractions depicted in (e).} 
\label{fig:NewTRG}
\end{center}
\end{figure}

\section{Application to tensor network renormalization}
\label{sect:E}
In this appendix we discuss the application of the proposed full environment truncation (FET) method towards tensor renormalization algorithms for the coarse-graining of path integrals and partition functions. Here the goal is to improve over the standard tensor renormalization group\cite{TRG1} (TRG) approach by removing internal correlations from within closed-loops of the network, similar to what was achieved with the tensor network renormalization\cite{TNR1,TNR2,TNR3,TNR4}  (TNR) approach and related algorithms\cite{TNRc1,TNRc2,TNRc3,TNRc4,TNRc5} which also remove internal correlations from closed loops.

We consider a square lattice tensor network with a 2-site unit cell, composed of 4-index tensors $A$ and $B$, as depicted in Fig. \ref{fig:NewTRG}(a). This network could be representative of the path integral of a $1D$ quantum system or the partition function of a $2D$ classical system, see for instance Ref. \onlinecite{TNR4}. An overview of an iteration of the proposed coarse-graining scheme is presented in Fig. \ref{fig:NewTRG}. The iteration begins with use of the SVD to decompose 4-index tensors into a product of 3-index tensors, identical to the standard TRG approach, where we retain at most $\tilde \chi$ singular values for each index. Then the FET scheme is applied to remove internal correlations within loops of 8 tensors, by sequentially truncating each of the four indices of dimension $\tilde \chi$ within the loop to a smaller dimension $\chi$, see also Fig. \ref{fig:NewTRG}(d). Finally, groups of tensors are contracted to form new 4-index tensors $A'$ and $B'$ as depicted in Fig. \ref{fig:NewTRG}(e), such that a coarser square lattice tensor network is obtained. These steps can can be iterated many times to generate a sequence of increasingly coarse-grained lattices. 

\begin{figure}[!t!hb]
\begin{center}
\includegraphics[width=6.0cm]{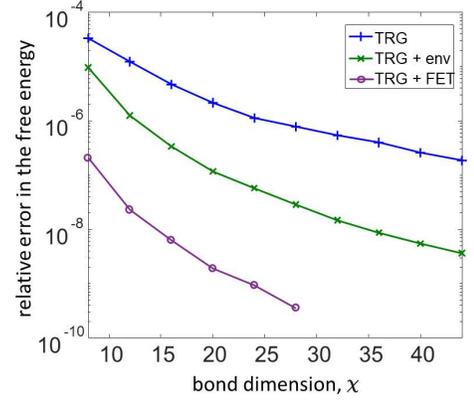}
\caption{Relative error in the free energy per site of the classical Ising model on a $2^{16}\textrm{-by-}2^{16}$ lattice of spins at critical temperature, comparing (i) tensor renormalization group\cite{TRG1} (TRG), (ii) tensor renormalization group with enlarged environment\cite{TRG5} (TRG + env) and (iii) tensor renormalization group that includes full environment truncations (TRG + FET).} 
\label{fig:Energy}
\end{center}
\end{figure}

This renormalization scheme is benchmarked through application to coarse-grain the $2D$ classical Ising model at critical temperature. We compare the scheme against standard TRG and an improved form of TRG\cite{TRG5} that takes a larger region of the environment into account in order to achieve greater accuracy. For each method, 32 coarse-graining steps are applied in order to reach a lattice size of $2^{16}\textrm{-by-}2^{16}$ classical Ising spins. The results for the (per-site) error in the free energy as a function of bond dimension $\chi$ are compared in Fig. \ref{fig:Energy}. It is seen that the TRG that includes the FET step significantly improves on standard TRG as well as the TRG with enlarged environment. With bond dimension $\chi = 28$, which required approximately 20 minutes computation time on a desktop PC, the TRG+FET scheme achieved a relative error in the free energy of $\delta f \approx 3 \times 10^{-10}$. The accuracy we achieve versus computation time appears to improve on the standard TNR\cite{TNR1} approach as well as the so-called loop-TNR approach\cite{TNRc1}, and to be comparable to the recently proposed GILT method\cite{TNRc4}. A key feature of the FET is that it is easily incorporated in any network geometry, similar to the GILT method, such that it could also be directly implemented, for instance, in higher dimensional networks.

\begin{thebibliography}{99}
\bibitem{real}
Note that we assume the tensors to be real-valued in order to avoid having to distinguish complex conjugates in the notation, although all of the results presented can easily be extended to complex-valued tensors.
\end{thebibliography}
\end{document}